\documentclass[aps,pra,twocolumn,showpacs]{revtex4}
\usepackage{graphicx}
\usepackage{epstopdf}
\usepackage{float}
\def\naive/{na\"{\i}ve}
\begin{document} 
\title{Efficient, Tightly-Confined Trapping of 226Ra} 
\author{R. H.~Parker,$^{1,2}$ M. R.~Dietrich,$^1$ K.~Bailey,$^1$ J. P.~Greene,$^1$ R. J.~Holt,$^1$ M. R.~Kalita,$^{1,3}$ W.~Korsch,$^3$ Z.-T.~Lu,$^{1,2}$ P.~Mueller,$^1$ T. P.~O'Connor,$^1$ J. Singh,$^1$ I. A.~Sulai,$^{1,2}$}\altaffiliation{Current address: Department of Physics, University of Wisconsin-Madison} \author{W. L.~Trimble$^1$}\altaffiliation{Current address: Institute for Genomics and Systems Biology, University of Chicago}
\affiliation{$^1$Physics Division, Argonne National Laboratory, Argonne, Illinois 60439, USA}
\affiliation{$^2$Department of Physics and Enrico Fermi Institute, The University of Chicago, Chicago, Illinois, 60637, USA}
\affiliation{$^3$Department of Physics and Astronomy, University of Kentucky, Lexington 40506, USA}

\date{\today}

\begin{abstract} We demonstrate a technique for transferring $^{226}$Ra atoms from a 3-dimensional magneto-optical-trap (MOT) into a standing wave optical dipole trap (ODT) in an adjacent chamber. The resulting small trapping volume (120~$\mu$m in diameter) allows for high control of the electric and magnetic fields applied to the atoms. The atoms are first transferred to a traveling-wave optical dipole trap, which is then translated 46 cm to a science chamber. The atoms are subsequently transferred into an orthogonal standing-wave ODT by application of a 1-dimensional MOT along the traveling-wave axis.  For each stage, transfer efficiencies exceeding 60\% are demonstrated. 
\end{abstract}


\maketitle

\section{Introduction}

There is great interest in using fundamental symmetries measurements to search for physics beyond the Standard Model. Many such measurements, including beta-neutrino angular correlation and triple-correlation coefficients~\cite{Oscar}, neutral atom atomic parity violation, and permanent electric dipole moments (EDMs) ~\cite{Review,Chupp} benefit when the species to be measured is confined in a tight, well-controlled, and environmentally-shielded trap. Optical dipole traps (ODTs) have been proposed as ideal traps for beta-decay asymmetry and neutral atom parity non-conservation experiments ~\cite{GRIMM,Sheng}, as well as EDM searches~\cite{Fortson}, due to the compact volume, low spin relaxation rate, and high degree of control of external electric and magnetic fields possible in ODTs. However, due to the low abundance of the isotopes used, it is often important that any technique have low atom losses. We demonstrate a technique for efficiently creating a high density of rare isotopes trapped in a standing-wave optical dipole trap (ODT), in a chamber where electric and magnetic fields can be both applied and controlled with great precision. 

Permanent electric dipole moments (EDMs) are sensitive probes of CP violation beyond the Standard Model prediction. Ongoing measurements studying the EDMs of neutrons, diamagnetic atoms, and paramagnetic atoms are each sensitive to different possible sources of CP violation~\cite{Review,Chupp}. The current best limit of the EDM of a diamagnetic atom was achieved with $^{199}$Hg~\cite{mercury}. $^{225}$Ra is another promising candidate for a diamagnetic atom EDM measurement, as it is expected to have an enhanced sensitivity to CP violating effects relative to $^{199}$Hg by a factor of a few hundred to a few thousand~\cite{schiff1,schiff2,schiff3}. However, for $^{225}$Ra (half-life = 15 days), the radioactivity limits reasonable atom number and the low vapor pressure prevents the use of a vapor cell, presenting significant experimental challenges.  Efforts toward $^{225}$Ra EDM searches are ongoing at Argonne National Laboratory~\cite{Proceedings} and Kernfysisch Versneller Instituut (KVI)~\cite{KVI}. 

\begin{figure}
\includegraphics[scale=0.7]{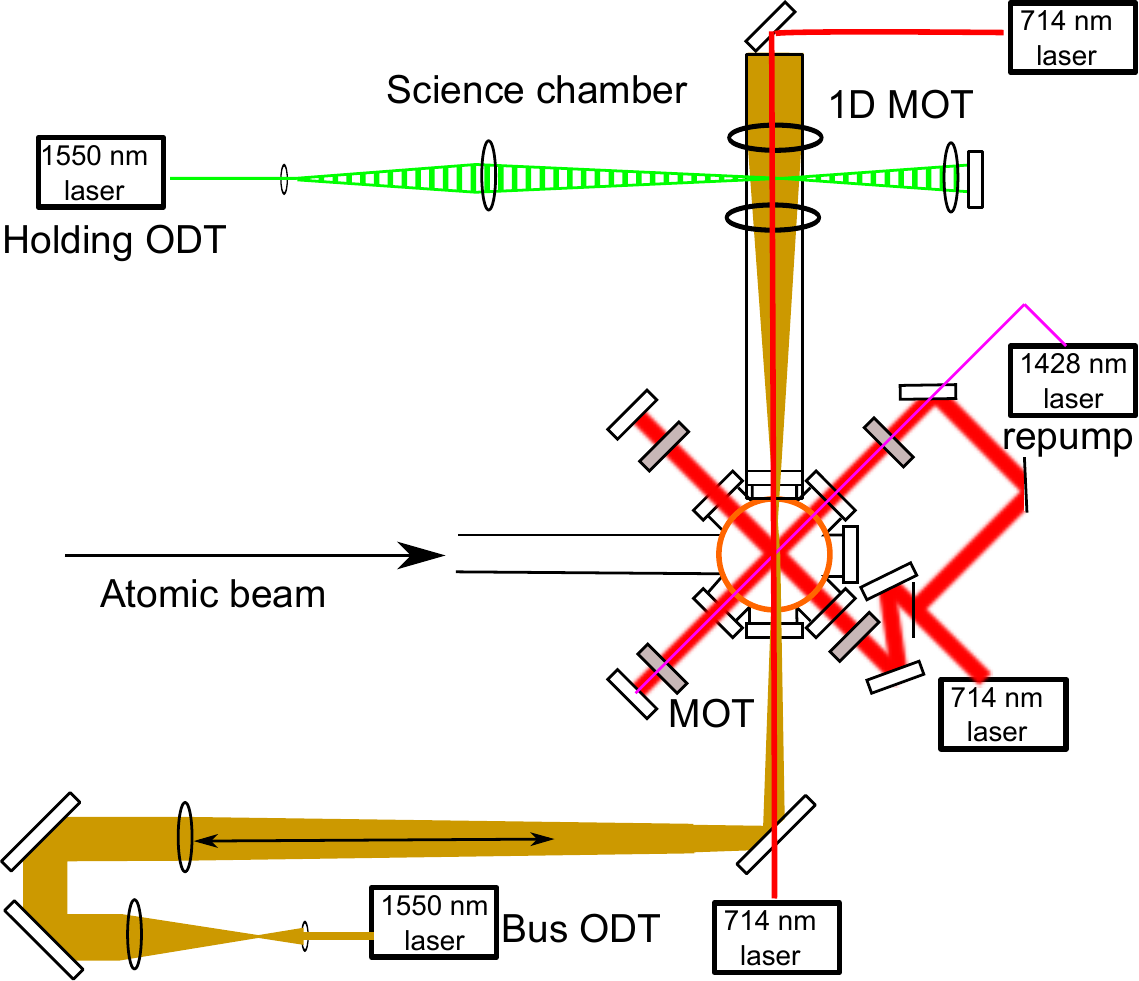}
\caption{(Color online) Diagram of the radium EDM apparatus. Atoms are loaded from an atomic beam into a 3D MOT and then transferred into an ODT. The ODT is translated to a science chamber in which they are transferred to a perpendicular `holding' ODT. The system is roughly 2~m x 2~m.}
\label{fig:Diagram}
\end{figure}

The radium EDM experiment underway at Argonne, shown in Figure~\ref{fig:Diagram}, employs a `conveyor belt' designed to transport atoms from an oven to a standing-wave ODT sandwiched between HV electrodes and surrounded by mu-metal magnetic shields. This is performed by first loading the atoms from a thermal atomic beam into a 3D MOT, then transferring the atoms into a traveling-wave ODT, next transporting the atoms from the magnetically-noisy MOT chamber to a magnetically-shielded science chamber, and finally transferring the atoms into a standing-wave ODT for the EDM measurement. As the EDM sensitivity scales as $1/\sqrt{N}$, where N is the number of atoms in the standing-wave, and the present supply of $^{225}$Ra is limited to less than about 250~ng per load, each stage must be as efficient as possible. Three stages are described in this paper---transfer of atoms from a 3D MOT to an ODT, transport of atoms in the ODT, and transfer of atoms from a traveling-wave bus ODT to an orthogonal standing-wave ODT. The use of a standing-wave ODT is necessary to reduce systematic uncertainties for the EDM measurement~\cite{Fortson,Chin}.

Two techniques have previously been used to transfer atoms between optical dipole traps. One has utilized atomic collisions to re-thermalize the atoms as the potential is altered~\cite{GRIMM}. The other has used shallow-angle optical dipole traps to increase the phase space overlap~\cite{Rolston}. These two techniques use adiabatic processes which require high atomic density. However neither technique is suited for the Argonne radium EDM experiment, in which the expected $^{225}$Ra density is too low to rely on atomic collisions, and the requirement of surrounding the atoms with electrodes and large, multiple-layered mu-metal magnetic shields makes the use of non-orthogonal ODTs inconvenient. Another method to cool the atoms is to use a MOT, but the limited optical access of the EDM experiment makes a 3D MOT difficult. Efficient transfer from a MOT chamber to a science chamber has also been demonstrated in a previous work~\cite{Weiss} by launching the atoms in a 1D lattice, followed by cooling in a 3D optical molasses through transparent (indium tin oxide-coated) electrodes. This however results in a large trap volume, which yields insufficient optical density for absorption imaging of scarce species such as $^{225}$Ra. We demonstrate that a suitable solution is to employ a MOT in only one dimension, relying on the restoring force of the ODT in the other two. 

\vspace*{0.5cm}
\section{Experimental Setup}
\subsection{MOT-to-ODT Transfer}
The setup for capturing radium atoms in a MOT from an atomic beam is described in~\cite{oldRa}. To review, the atomic beam is emitted by an oven and passes through a transverse cooling stage and Zeeman slower before loading into the MOT. All three laser cooling steps are operated on the 714nm $^{1}$S$_{0}$$\rightarrow$$^{3}$P$_{1}$ intercombination transition ($\Gamma=2\pi \times$ 380 kHz~\cite{Scielzo}, I$_{sat}$ = 140 $\mu$W/cm$^{2}$). For testing purposes, the experiment uses the spinless isotope $^{226}$Ra, which has a half-life of 1600 years. The MOT lifetime is 20s, typically resulting in a MOT of 16000 $^{226}$Ra atoms for a loading time of 10s. As the EDM measurement is based on spin-precession in a magnetic field, the experiment will use $^{225}$Ra (15-day half-life) for the actual EDM measurement, which is more scarce, but has a nuclear spin of 1/2. 


The 3D MOT part of each experimental cycle is divided into a loading phase (for transferring atoms from the atomic beam to the MOT), a probing phase (for diagnostic imaging of the MOT), and a cooling phase (for transferring the atoms from the MOT to the ODT). This division allows nearly independent optimization of each stage. During the loading phase the laser intensity is 1.7~mW/cm$^{2}$, the detuning is 2.8~MHz to the red of resonance, and the magnetic field gradient is 1~G/cm. During the probing phase the laser intensity is decreased to 500~$\mu$W/cm$^{2}$, the detuning is decreased to 2~MHz, and the magnetic field gradient is increased to 2.5~G/cm. During the cooling phase, the laser intensity is further decreased to 40~$\mu$W/cm$^{2}$ while the detuning from resonance is decreased to 1.1~MHz, with no change to the magnetic field gradient. We obtain an atom temperature of 40~$\pm$ 15~$\mu$K, measured by both a time-of-flight technique and a release-and-recapture technique~\cite{Tuchendler}. 

During the cooling phase the atoms from the MOT are transferred into a traveling-wave ODT (hereafter referred to as the `bus ODT') generated from a 1550~nm 50~W multimode, unpolarized fiber laser (IPG ELR-50-1550). This laser is expanded to fill a 10~cm diameter 2~m focal length lens, such that the focus of the lens overlaps the MOT with a waist diameter of 100~$\mu$m. 1550~nm was chosen as it is the predicted `magic wavelength' for the intercombination transition used in the MOT~\cite{Dzuba1,Dzuba2}, and is thus expected to yield high transfer efficiencies. The bus ODT produces a trap depth of 540~$\mu$K. The lifetime of atoms in the bus ODT depends on the pressure; for a typical pressure of 4$\times$10$^{-10}$~Torr in the MOT chamber, the ODT lifetime is about 8~seconds, a factor of 3 less than the MOT lifetime. The large atomic mass of radium and the low longitudinal trap frequency (5.5~Hz) requires alignment of the ODT with respect to gravity to better than 10~mrad.

\subsection{ODT Transport}

The next step is to transport the atoms from the MOT chamber to the science chamber in the bus ODT, 46~cm away~\cite{Dzuba1,Dzuba2}. As the cooling phase is now complete, the 714~nm light is turned off so that the only trap is due to the ODT potential. The 2~m focal length lens is mounted on an air-bearing magnetically-actuated translation stage (Aerotech ABL2000), which moves horizontally from the center of the MOT chamber to the center of a science chamber. Moving this lens translates the bus ODT focus longitudinally, which causes the trapped radium atoms to move along with it. The main loss mechanisms are from the motion of the trap itself and background gas collisions. Background gas collisions can be reduced with improvements to the vacuum system; to optimize the motion profile the loss due to the motion itself must be measured.

Transport efficiency is measured by comparing the number of atoms surviving after a round-trip of a given distance to the number of atoms in the MOT before the motion began; by applying a square root to this ratio, we learn what fraction of atoms were lost in each part of the trip (both motions follow the same profile). This uses the assumption that there are no additional losses due to the turn-around at the far end of the travel, which is supported by simulation. This technique involves one imaging method for both measurements, and is thus insensitive to systematic effects associated with using two different imaging schemes. Survival can be measured by dividing the number of atoms that survive the round-trip travel by the number of atoms that remain in the ODT if no motion is attempted, keeping the imaging time the same. This cancels out the loss due to the ODT lifetime, and gives the loss due solely to the motion itself. Because the ODT lifetime is different in the two chambers, transport distances were limited to 300~mm (rather than 460~mm) to keep the ODT lifetime the same when characterizing transport efficiency.

\begin{figure}
\includegraphics[scale=0.28]{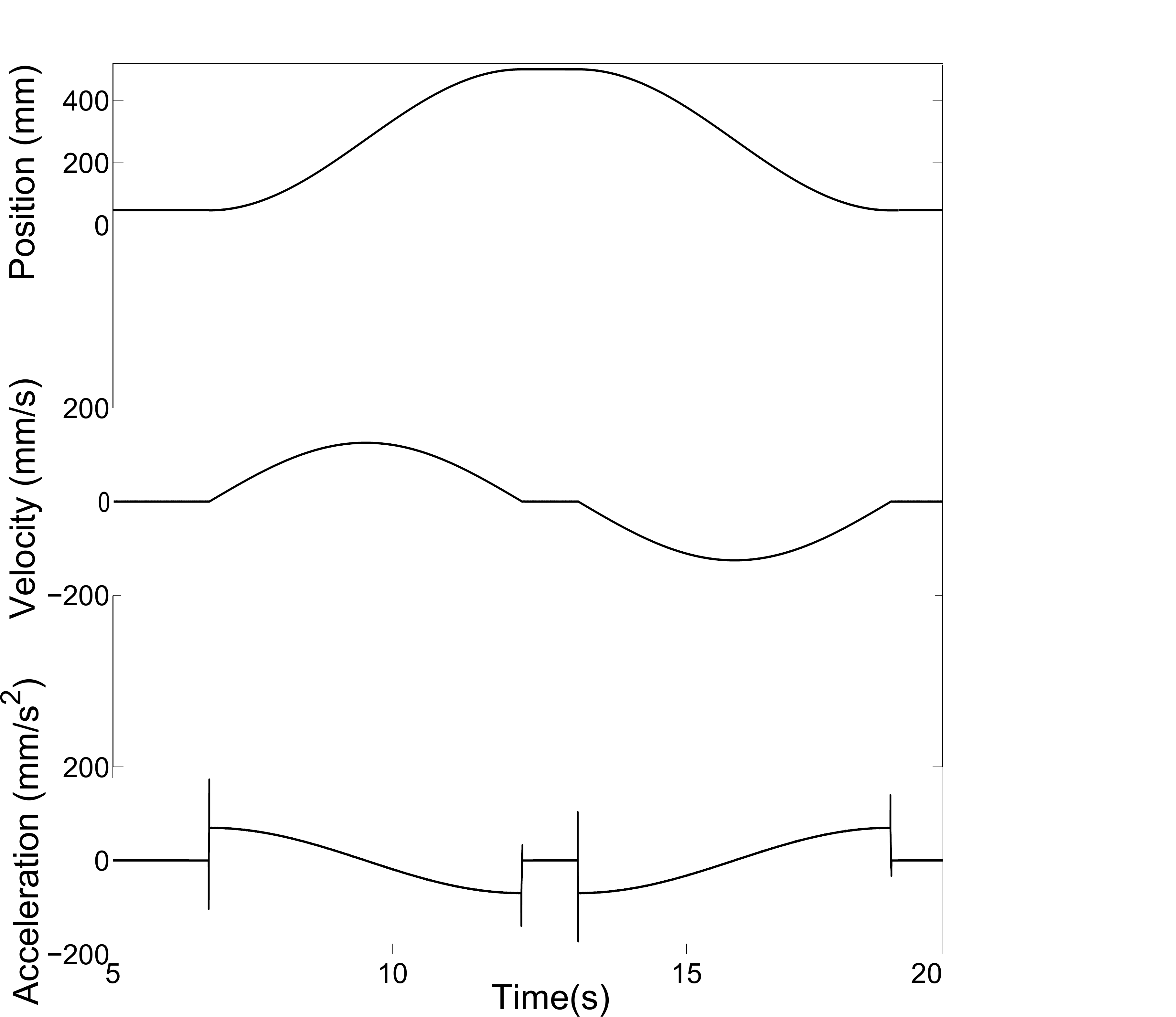}
\caption{Example data from the translation stage, for a transport distance of 460~mm in 5.7~seconds. The top, middle, and bottom graphs are position vs time, velocity vs time, and acceleration vs time for a round trip. Position data is obtained from a linear encoder on the stage; differentiation yields the velocity and acceleration profiles.}
\label{fig:Profile}
\end{figure}

Various functional forms for the ODT position vs time were tested: sinusoidal, triangular, parabolic, and `minimal jerk' (a linear combination of sinusoidal and parabolic intended to minimize the maximum derivative of acceleration vs time, at the expense of increased transport time). The motion profile that optimized atom number transported was found to be sinusoidal, shown in Figure~\ref{fig:Profile}.

\subsection{ODT-to-ODT Transfer}

The 1D MOT in the science chamber is formed by a pair of quadrupole coils aligned along the bus ODT axis and three sets of trim coils for the three orthogonal axes. These coils produce a magnetic field gradient of 0.75~G/cm along the bus ODT axis, with a magnetic field zero near the overlap of the two ODT's. The intensity used in the 1D MOT is typically 10~$\mu$W/cm$^{2}$. Two opposite circularly-polarized 714~nm beams are aligned longitudinal to the bus ODT for creating the 1D MOT (called the `longitudinal' beams in this paper). The longitudinal beams come independently from the two directions rather than being a single retroreflected beam, which allow the powers in the beams, and thus the position of the 1D MOT, to be tuned. The atoms in the science chamber are imaged by 1D MOT fluorescence with a CCD camera (Andor Luca$^{EM}$ R).  The pressure in the science chamber is typically 6$\times$10$^{-11}$~Torr, giving a lifetime in the holding ODT of roughly 14~seconds.

A standing-wave ODT (the `holding ODT') produced from a retroreflected 1550~nm 10~W single-mode linearly polarized fiber laser (IPG ELR-30-1550-LP-SF) is aligned such that its focus overlaps with the focus of the bus ODT after the 46~cm travel. The waists of the two beams are the same. The standing wave trap thus has a 1D lattice depth comparable to the depth of the traveling-wave bus ODT, about 430~$\mu$K. A 200~ms pulse of longitudinal 714~nm 1D MOT light compresses the atoms and loads them into the overlap of the two ODTs. The bus ODT and longitudinal beam powers are then turned off for 100~ms, long enough for atoms not loaded into the holding ODT to fall under gravity. Thus any atoms that remain have been loaded into the holding ODT. 

\begin{figure}
\includegraphics[scale=0.2]{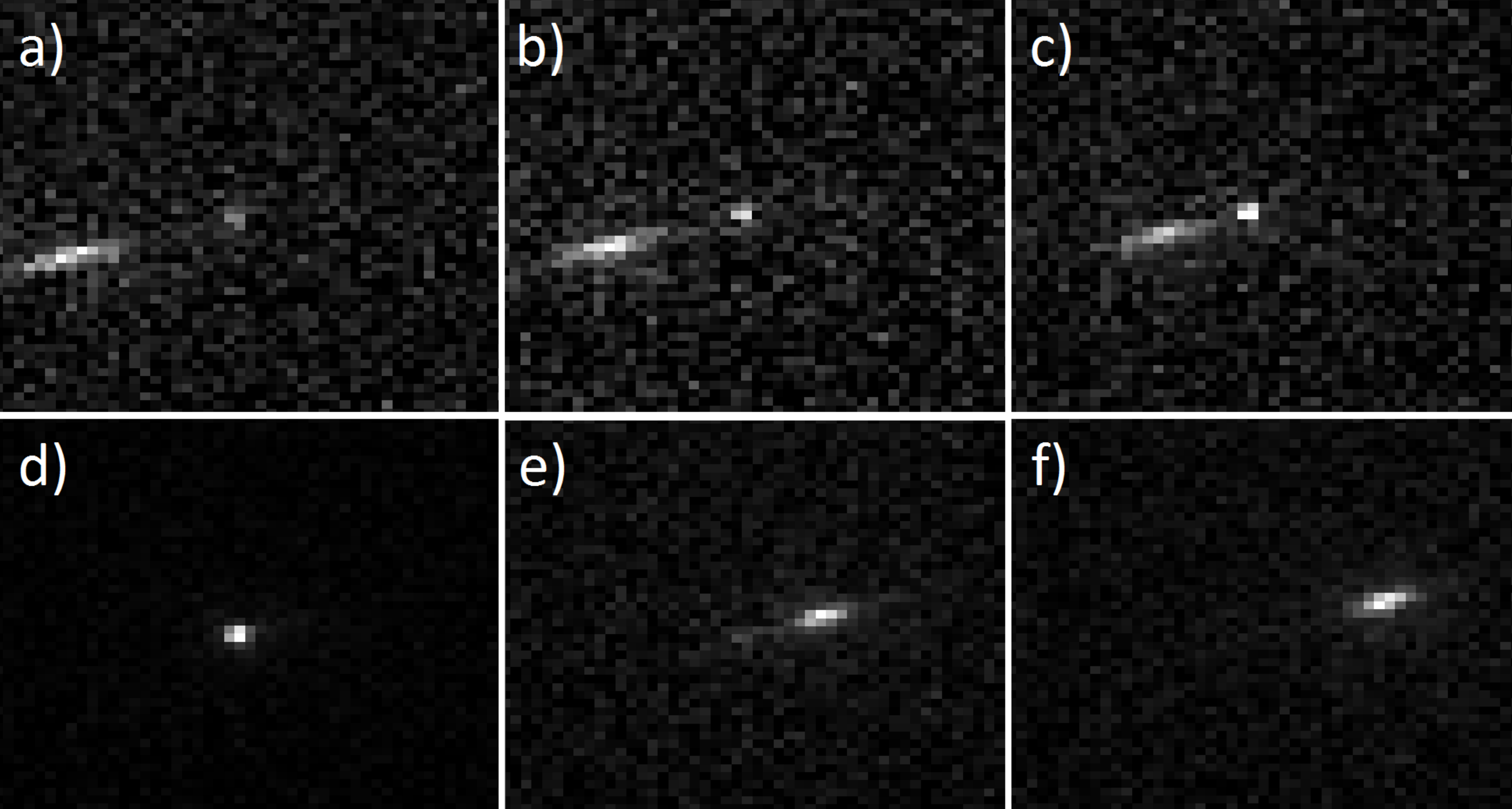}
\caption{Adjusting the position of the 1D MOT relative to the overlap of the two ODT's by changing the balance of the 1D MOT beams. From a) to f) the 1D MOT is swept through the overlap of the two ODT's. Each image is a single shot of atoms. Fluorescence generally occurs most strongly in two places; the center of the 1D MOT (the ellipse) and the deep potential well (the dot) created by the two ODTs. All three are aligned in d); this is the position for optimum transfer efficiency. Each image is 2 mm wide.}
\label{fig:1DMOT}
\end{figure}

The longitudinal light pulse generates a 1-dimensional MOT. The atoms are confined in the longitudinal axis primarily by the non-conservative MOT forces, and in the transverse axes only by the conservative ODT potential. This creates a tightly compressed atom cloud. By tuning the trim coils or longitudinal balance we can adjust the 1D MOT location, as in Figure~\ref{fig:1DMOT}. By aligning this MOT with the overlap of the two ODTs, it is possible to transfer the atoms from the bus ODT to the holding ODT. 

We use two independent measures of the ODT-to-ODT transfer efficiency, to avoid systematic effects. In the first, we compare the fluoresence in the science chamber with and without the dropping of the bus ODT. In the second, we return any untransferred atoms to the MOT chamber, where their number can be more accurately measured. Each measure of transfer efficiency is the ratio of the fluorescence from two timing sequences, one with transfer and one without. 

For the first measure of transfer efficiency we compare the result of two sequences. In the first, once the atoms arrive in the science chamber there is a delay of 400~ms followed by 200~ms with longitudinal beams on. The bus ODT and the longitudinal beams are then turned off for 100~ms, after which the longitudinal beams are turned back on and an image is taken. In the second sequence, once the atoms arrive in the science chamber there is a delay of 700~ms, followed by a longitudinal pulse and imaging. This allows us to identify the ratio of the first sequence to the second as the transfer efficiency. 

The second measure of transfer efficiency utilizes fluorescence in the MOT chamber. In the first sequence of this scheme, after the atoms are loaded into the holding ODT, the bus ODT is transported back to the MOT chamber, where an image is taken. The second sequence is simply a wait of 2~seconds after the atoms arrive in the science chamber, followed by a trip back to the MOT chamber and an image taken. The ratio of the two thus gives the fraction of atoms not transferred to the holding ODT, which in turn gives the transfer fraction. The two definitions yield efficiencies that agree to within 1-sigma.

\vspace*{0.5cm}
\section{Results}
The parameters used in the cooling phase result in MOT-to-ODT transfer efficiencies of 75~$\pm$ 5\% under routine operating conditions, where the error given is the statistical uncertainty per shot. The dominant systematic in this measurement is a change in the power in the MOT light during the probe phase caused by thermal drift in the double-pass AOM used to tune the frequency of the MOT light, which results in the pre-ODT-drop and post-ODT-drop images having two different MOT powers; however, this can be tuned to be much less than the statistical fluctuations of the MOT. The MOT statistics are limited by atom shot noise and frequency instability; fluctuations due to chages in the background and the position of the MOT are negligible. Efficiencies as high as 96\% have been achieved, with regular maintanence of the apparatus. This transfer efficiency is attributed to the low density of atoms, the use of a far-off-resonant magic-wavelength ODT, and narrow-line cooling. 

\begin{figure}
\includegraphics[scale=.75]{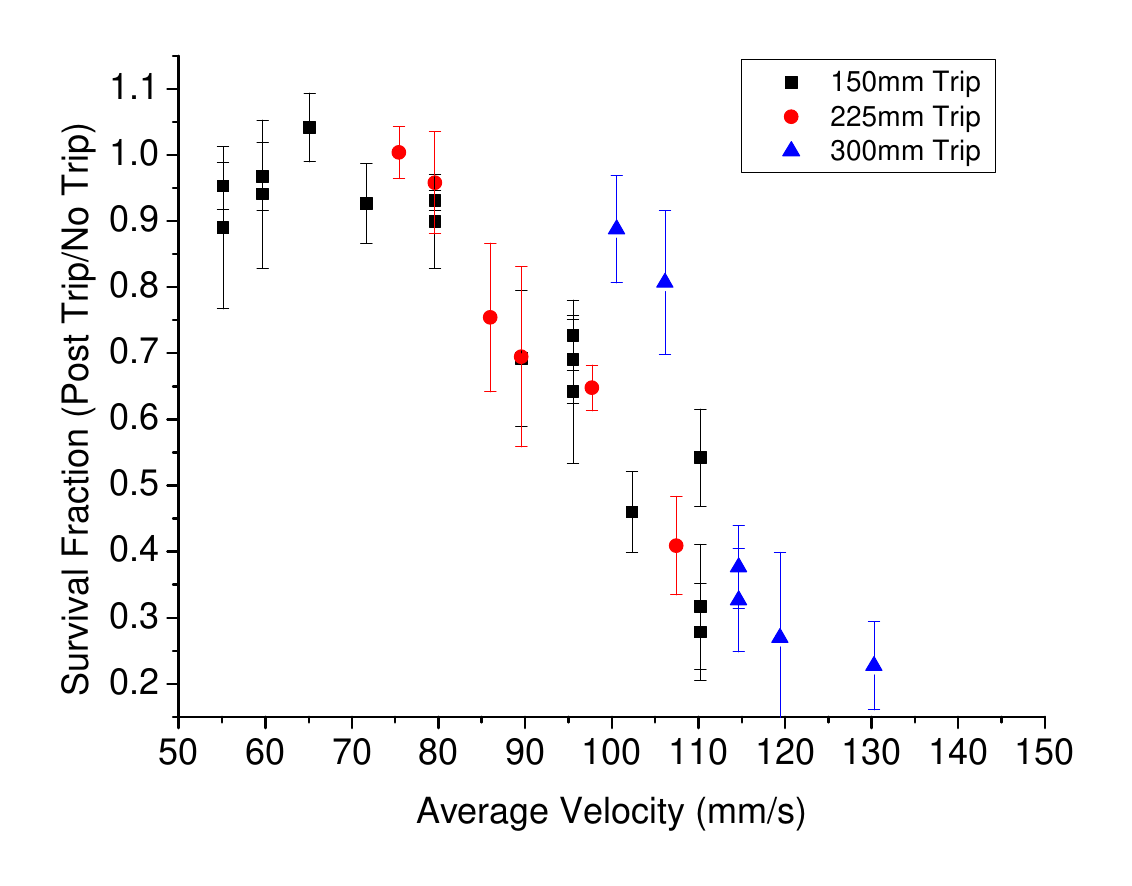}
\caption{(Color online) Survival for three transport distances as a function of average velocity, with correction applied for the ODT lifetime. The distances given are half the round-trip distance. Typically 5 runs were performed per data point. All runs were done using a sinusoidal motion profile, an example of which is shown in Figure~\ref{fig:Profile}.}
\label{fig:Transport2}
\end{figure}

\begin{figure}
\includegraphics[scale=1.6]{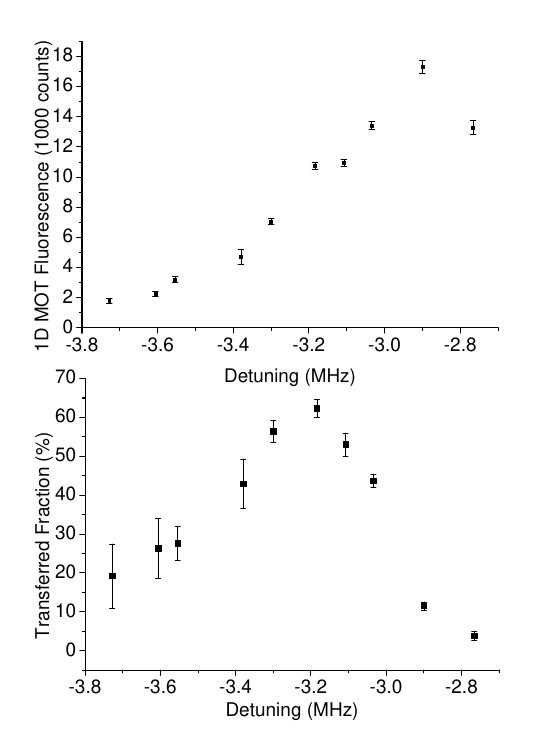}
\caption{Graph a) shows the fluorescence from the 1D MOT as a function of laser cooling detuning, relative to the $^{1}$S$_{0}$$\rightarrow$$^{3}$P$_{1}$ resonance. Graph b) shows the resulting ODT-to-ODT transfer efficiency, using the first definition described in the text. The plots show 1-sigma statistical error bars.}
\label{fig:Handoff}
\end{figure}

Figure~\ref{fig:Transport2} shows lifetime-corrected round-trip survival fractions for three different transport distances, as a function of average velocity. It is clear that for sufficiently slow motions, all the losses are due to background-gas collisions; the losses due to the motion itself are negligible. As the motion speed increases the losses due to the ODT lifetime decrease but the losses due to the motion increase--therefore for a given transport distance and pressure there is an optimum transport time and a maximum possible transport efficiency. For the 46~cm transport distance to the science chamber for pressures of 10$^{-11}$--10$^{-10}$~Torr, the one-way transport efficiency is measured to be 60\% (not correcting for lifetime losses) for routine operation, and the optimum transport time is 5.7~seconds. As the ODT transport is measured using a ratio of MOT images, it is subject to the same statistical and systematic effects as the MOT-to-ODT transfer; the repeatability of the translation stage lens has been found to be much better than the intrinsic fluctuations of the MOT size.

In characterizing the 1D MOT as a tool for ODT-to-ODT transfer it is useful to determine its sensitivity to experimental parameters. Particularly important is the sensitivity of the transfer efficiency to the frequency of the cooling light, as this places requirements on the stability of the laser lock and the laser linewidth. The results are shown in Figure~\ref{fig:Handoff}, with the frequency defined relative to the resonance observed in the 3D MOT. Transfer efficiencies exceeding 60\% have been demonstrated. The frequency width is consistent with the $^{1}$S$_{0}$$\rightarrow$$^{3}$P$_{1}$ linewidth of $\Gamma=2\pi \times$ 380~kHz. Due to the low depth of the trap in single-pass, no transfer is measured without the retroreflection of the holding ODT. The detuning for maximum 1D MOT fluorescence is 2.9~MHz to the red of resonance, while the detuning for optimum transfer into the holding ODT is 3.2~MHz. The large detuning relative to the radium $^{1}$S$_{0}$$\rightarrow$$^{3}$P$_{1}$ resonance is attributed to the residual AC Stark shift of the two ODTs and stray magnetic fields. The dominant systematic for this measurement is expected to be the residual AC Stark shift of the atoms in the bus ODT, as the atoms have slightly different scattering rates before and after they are loaded into the holding ODT. The significance of this systematic was determined by comparing the two different measures of transfer as described in the previous section; the second measure relies on fluorescence in the 3D MOT and is thus insensitive to the AC Stark shift systematic. As the agreement of the two methods is within 1-sigma, the AC Stark shift systematic does not limit the measurement. 

\vspace*{0.5cm}
\section{Discussion}
Before discussing the results of ODT-to-ODT transfer with a 1D MOT, it is worth explaining why simple longitudinal cooling along the bus ODT axis with no quadrupole field does not work. Any technique for efficiently transferring atoms between two traps requires that the rate at which atoms are loaded into the second trap be significantly greater than the rate at which atoms are lost through the loading process. For example, MOT-to-ODT transfer can be made very efficient because the time required for loading is roughly 200 ms, while the lifetimes of the atoms in the MOT and ODT are 20~s and 8~s respectively. Qualitatively, it can be seen that 1D longitudinal cooling alone produces an optical molasses in which the atoms have a diffusion time of several seconds in the longitudinal direction---thus the loading rate is very low. However, the atoms are scattering into the transverse dimensions with no cooling along those axes; the atoms are thus lost rapidly through boiling out of the (conservative) ODT. Atoms can also be pumped to metastable states which may not be trapped in the ODT.  The lifetime of the atoms under illumination has been measured to be about 400~ms. 

For transfer using a 1D MOT, on the other hand, the loading time is roughly 200~ms, the same as for the 3D MOT. This yields maximum transfer efficiencies of about 60\%; less than the $\>$96\% possible with MOT-to-ODT transfer, but still sufficient for many applications. We thus routinely get efficiencies exceeding 25\% for the full MOT-to-holding ODT transfer process, including all three steps. This can be compared to the 78\% transfer between two 3D MOTs that has been demonstrated by launching the atoms through two atomic funnels~\cite{Swanson}, which is unfortunately not applicable for an EDM experiment. 

For experiments in which a residual magnetic field does not cause limiting systematic effects, the procedure as described above is sufficient. However, for experiments sensitive to net magnetic fields, such as an EDM measurement, the DC magnetic quadrupole of the 1D MOT may be undesirable. To solve this potential problem, the 1D MOT can be made to operate with AC instead of DC magnetic fields, in which the polarization of light and sign of magnetic field are synchronously alternated at high frequency. This results in a zero time-averaged magnetic field while still providing cooling and confinement. Such an AC MOT has been shown to have a lifetime comparable to that of the DC version~\cite{ac}.

Realization of efficient transfer between ODTs was an essential step for the Argonne $^{225}$Ra EDM experiment. The 1D MOT technique demonstrated in this paper provides high transfer efficiency and is widely applicable for many other atomic species.

\vspace*{0.5cm}
\section{Acknowledgements}
We would like to thank C. Chin, V. A. Dzuba, and S. L. Rolston for helpful discussions. This work is supported by DOE Office of Nuclear Physics, Contract No. DE-AC02-06CH11357.

\end{document}